\documentclass{article}
\usepackage{amssymb}
\usepackage{amsmath}
\usepackage{amsfonts}
\usepackage{graphics,graphicx,epsfig}
\usepackage{color}
\usepackage{float}
\usepackage{hyperref}
\usepackage{caption}
\usepackage{multicol}
\usepackage{cancel}
\usepackage{soul}
\usepackage{enumitem}
\usepackage{mathrsfs}
\usepackage{rotating}
\usepackage[normalem]{ulem}
\usepackage{booktabs} 

\begin{document}

\title{ADVANCED BIOMARKER ANALYSIS FOR EARLY ALZHEIMER'S DETECTION: A 3-CLASS CLASSIFICATION APPROACH
}
\author{Victor Miguel Sierra Marquina$^{1}$, Maria del Carmen Pardo$^{2,3},$ Alba Maria Franco-Pereira$^{2,3,*}$\\
\small%
$^{1}$Complutense University of Madrid,
28040-Madrid, Spain\\
\small%
$^{2}$Department of Statistics and O.R., Complutense University of Madrid,
28040-Madrid, Spain\\
\small%
$^{3}$Instituto de Matem\'{a}tica Interdisciplinar, Complutense University
of Madrid, 28040-Madrid, Spain\\
\small *corresponding author (albfranc@ucm.es)\\
}
\date{}
\maketitle

\begin{abstract}

The receiver operating characteristic (ROC) curve is an important tool for the discrimination of two populations. However, in many settings, the diagnostic decision is not limited to a binary choice. ROC surfaces are considered as a natural generalization of ROC curves in three-class diagnostic problems and the Volume Under the ROC Surface (VUS) was proposed as an index for the assessment of the diagnostic accuracy of the marker under consideration. In this paper, we propose an overlap measure (OVL) in the case of three-class diagnostic problems. Specifically, parametric and non-parametric approaches for the estimation of OVL are introduced. We evaluate this measure through simulations and compare it with the well-known measure given by VUS. Furthermore, our proposal is applied to the clinical diagnosis of early stage Alzheimer’s disease.

\end{abstract}


\section{Introduction}\label{Introduction}

Diagnostic testing is an extremely important aspect of medical care. A common way to evaluate biomarkers for disease diagnosis involve the use of ROC curves, and the Area Under the Curve (AUC) as measures of test accuracy (\hyperref[Nakas et al. (2023)]{Nakas et al., 2023}). Biomarkers are crucial for disease prevention and detection, and their evaluation involves analyzing sensitivity (probability of correctly diagnosing diseased patients) and specificity (probability of correctly identifying non-diseased patients). Diagnostic tests vary in scale and distribution, so measures that summarize and evaluate biomarker discrimination ability are needed. Many resources and R software packages address this for binary markers (healthy vs. diseased). However, many diseases have an intermediate population in an early or transitional stage that requires proper identification for early intervention and increased survival probability. The lack of research and statistical tools for three-population scenarios can limit disease prevention and early detection efforts. For instance, mild cognitive impairment (MCI) and/or early stage Alzheimer’s disease is a transition stage between the cognitive changes of normal aging and the more serious problems caused by Alzheimer’s disease (AD). AD is a progressive neurodegenerative disorder that leads to the death of brain cells that cannot be replaced once lost. The transitional stage in the AD process is critical to detect not only because it is indicative of more serious disease development in the future but also because this is the time for family caregivers to discuss care options and family future planning while the patient can still take part in making decisions (\hyperref[Xiong et al., 2006]{Xiong et al., 2006}).

When it is critical to classify individuals into three groups (i.e., normal, early disease stage, fully developed disease stage), \hyperref[Scurfield (1996)]{Scurfield (1996)} introduced the ROC surface as a natural generalization of the ROC curve. \hyperref[Mossman (1999)]{Mossman (1999)} proposed a polyhedral ROC surface and the volume under the surface as a summary measure of the diagnostic accuracy. \hyperref[Dreiseitl et al.(2000)]{Dreiseitl et al.(2000)} further provided a non-parametric estimate and the associated variance for the volume under a polyhedral ROC surface with probabilistic rating data. The ROC surface for multi-class diagnostic problems, in a non-parametric context, was proposed by \hyperref[Nakas and Yiannoutsos (2004)]{Nakas and Yiannoutsos (2004)}. \hyperref[He and Frey (2008)]{He and Frey (2008)} also discussed the nonparametric estimation of a single VUS. Both parametric and nonparametric inferences on the VUS have been developed (\hyperref[Xiong et al., 2006]{Xiong et al., 2006}; \hyperref[Li y Zhou (2009)]{Li and Zhou, 2009}; \hyperref[Inácio, Turkman, Nakas, and Alonzo, 2011]{Inácio et al., 2011}). \hyperref[Kang-Tian]{Kang and Tian (2013)} offered an extensive study comparing possible parametric and non-parametric approaches for the estimation of VUS in terms of bias and root mean square error. 

The VUS is an extension of the total area under the ROC curve (AUC). However, the drawbacks of the AUC for the classification in two categories are known and many other alternative summary indices have been proposed (\hyperref[reffra]{Franco Pereira et al., 2020}). Recently, overlapping coefficients (OVL) are widely used in many fields, for example authors in \hyperref[refmi]{Mizuno et al. (2005)} compared pharmacokinetic parameters of the same drug for two ethnically different regions; the niche overlap of two species, which uses up to $n$ different resources, is addressed by \hyperref[refmu]{Mueller and Altenberg (1985)}; the proportion of devices that have a similar range of failure time has been analyzed by \hyperref[refal]{Al-Saleh and Samawi (2007)}. \hyperref[Pardo and Franco-Pereira (2024)]{Pardo and Franco-Pereira (2024)} explored the advantages of the overlap measures over the ROC summary indices to assess the accuracy of a medical diagnostic test in the binary case. The overlap measures have some desirable properties such as invariance to monotone increasing transformations as the ROC summary indices. AUC does not capture shape differences whereas OVL has the advantage of being able to detect differences in both location ans shape between the underlying distributions of diseased and healthy samples. Moreover, AUC yields improper values when it is 0.5 or lower, which occurs when the mean score for diseased individuals is less than the mean score for healthy individuals. Therefore, the overlap measures have been proved to be better than the ROC summary indices for assessing diagnostic tests, not only because they perform well when the ROC summary indices fail but also because they are comparable or even superior when the latter are effective. \hyperref[Franco Pereira et al.]{Franco-Pereira et al. (2021)} addressed the need for robust methods to construct confidence intervals for OVL when dealing with two-class problem by proposing both parametric and non-parametric approaches and recommended different  cutoff values for classifying differentiation quality. 

\[
\begin{cases}
OVL = 1 & \text{No differentiation}, \\
1 > OVL > 0.75 & \text{Poor differentiation}, \\
0.75 > OVL > 0.55 & \text{Good differentiation}, \\
0.55 > OVL > 0.35 & \text{Very good differentiation}, \\
0.35 > OVL & \text{Excellent differentiation}.
\end{cases}
\]

Given the advantages of OVL in the two-class case, the aim of this study is to extend and investigate the OVL measure to the three-class classification problem. In this article, we propose an overlap measure to discriminate among three diagnostic groups. Several parametric and non-parametric estimation approaches are considered for this measure. Next, we apply our methodology to estimate the diagnostic accuracy of early stage AD for a real data set using a couple of biomarkers. Finally, an extensive simulation study comparing OVL and VUS in terms of power is conducted. 

\section{VUS and OVL measures. Estimation Methods}\label{Section 2}
To formally define the classification problem with three classes, let's assume $n_1$ measurements from the first class, denoted by $X_1$, following a distribution function $F_1$, i.e., $ X_1 \sim F_1 $, $n_2$ measurements from the second class, $ X_2 \sim F_2 $, and $ n_3 $ measurements from the third class, $X_3 \sim F_3$. Specifically, we can assume that $X_1<X_2<X_3$. 

We can define a decision rule $\mathscr{D}$, which assigns each element to one of the three classes, by establishing two thresholds, $c_1<c_2$, i.e., if $X$ is a new element to be classified, we assign $X$ to

\begin{equation*}
\mathscr{D} = 
\begin{cases}
  X_1 & \text{if } X \leq c_1, \\
  X_2 & \text{if } c_1 < X \leq c_2, \\
  X_3 & \text{if } X > c_2.
\end{cases}
\end{equation*}

This classification leads to three true positive fractions (TPF) and six false positive fractions (FPF). We define $TPF_1=P(X_1 \leq c_1)$, $TPF_2=P(c_1<X_2 \leq c_2)$, and $TPF_3=P(X_3>c_2)$.

If $p_1=TPF_1$ and $p_3=TPF_3$, the functional form of the ROC surface could be expressed as follows:
\begin{equation}\label{ROC def}
\text{ROC}_s(p_1,p_3)=
\begin{cases}
  F_2\left(F_{3}^{-1}(1-p_3)\right)-F_2\left(F_{1}^{-1}(p_1)\right) & \text{if} \quad F_{1}^{-1}(p_1) \leq F_{3}^{-1}(1-p_3),   \\
  0 & \text{otherwise}. \\
\end{cases}
\end{equation}

As a measure of classifier accuracy with 3 classes, VUS is defined:
\begin{equation}
VUS = \int_{0}^{1} \int_{0}^{1-F_3\left(F_{1}^{-1}(p_1)\right)} \text{ROC}_s(p_1,p_3) \, dp_3 \, dp_1.
\label{VUS}
\end{equation}
The minimum VUS for a three-class classifier is $\frac{1}{6}$, which is the volume of the tetrahedron obtained when $X_1$, $X_2$ and $X_3$ are identically distributed. Its maximum value is one and corresponds to the case in which the three classes are perfectly discriminated in the correct order.

An alternative to the VUS to assess the ability of a marker to differentiate three classes are the overlap measures. The most commonly used was introduced by \hyperref[Weitzman (1970)]{Weitzman (1970)}. It is defined as 
\begin{equation}
OVL=\int_{-\infty}^{+\infty} \min \left\{f_1(x), f_2(x), f_3(x) \right\} \, dx,
\label{OVL}
\end{equation}
where $f_i(x)$ is the density function associated to $X_i$, $i=1,2,3$.
Clearly, in general, $0 \leq OVL \leq 1$. OVL measures the similarity between distributions by quantifying the overlapping area of their respective density functions. Its intuitive interpretation and the ability to visualize overlap through sample histograms make it particularly attractive to applied researchers. This measure has been also the base for $k$-sample tests (\hyperref[refmar]{Martinez-Camblor et al., 2008}).

\subsection{Estimation Methods}

\subsubsection{Parametric estimation}

When biomarker levels follow normal distributions, i.e., $X_i \sim \mathcal{N}(\mu_i,\sigma_i)$, the volume under the trinormal ROC curve after substituting the unknown parameters by the maximum likelihood estimators (MLE) obtained from samples of the three distributions is:

$$\widehat{VUS}_N = \int_{-\infty}^{+\infty} \Phi(as-b) \Phi( -cs + d) \phi(s) \ ds,$$
where $a=\frac{\widehat{\sigma_2}}{\widehat{\sigma_1}}$, $b=\frac{\widehat{\mu_1}-\widehat{\mu_2}}{\widehat{\sigma_1}}$, $c=\frac{\widehat{\sigma_2}}{\widehat{\sigma_3}}$, $d=\frac{\widehat{\mu_3}-\widehat{\mu_2}}{\widehat{\sigma_3}}$, $\Phi(\cdot)$ denotes the standard normal distribution function, and $\phi(\cdot)$ its density function. 

The corresponding OVL measure can be written as 
$$\widehat{OVL}_N = \int_{-\infty}^{+\infty} \min \left\{ \frac{1}{\widehat{\sigma_1}}\phi\left(\frac{x-\widehat{\mu_1}}{\widehat{\sigma_1}}\right), \frac{1}{\widehat{\sigma_2}}\phi\left(\frac{x-\widehat{\mu_2}}{\widehat{\sigma_2}}\right), \frac{1}{\widehat{\sigma_3}}\phi\left(\frac{x-\widehat{\mu_3}}{\widehat{\sigma_3}}\right) \right\} \ dx.$$

The assumption that biomarkers are normally distributed can be very limiting, resulting in inaccurate outcomes if the assumption is significantly violated. A parametric approach based on the Box–Cox transformation to normality often works well in this context (\hyperref[Bantis-Nakas]{Bantis et al., 2017} and \hyperref[Franco Pereira et al.]{Franco-Pereira et al., 2021}). It consists of assuming that a common, usually unknown, monotone transformation results in the three distributions being normal.

\subsubsection{Box-Cox transformation}
\hyperref[Box-Cox]{Box and Cox (1964)} proposed a parametric family of power transformations that enables the derivation of variables with a more stable variance and cumulative distribution function closer to the normal distribution.

Let $X_{11},X_{12},...,$ $X_{1n_{1}}$, $X_{21},.X_{22},..,$ $X_{2n_{2}}$ and $X_{31},.X_{32},..,$ $X_{3n_{3}}$
denote three random samples of sizes $n_{1}$, $n_2$ and $n_{3}$. For each value of the parameter $\lambda$, the transformation from $X_{ij}$ to $X_{ij}^{(\lambda)}$ is considered ($i=1,2,3$), where $X_{ij}>0$ and the expression for $X_{ij}^{(\lambda)}$ is given by

\begin{equation*}
X_{ij}^{(\lambda)}=
\begin{cases}
  \frac{X_{ij}^{\lambda}-1}{\lambda} & \text{if} \quad \lambda \neq 0,  \\
  \log({X_{ij}}) & \text{if} \quad \lambda = 0. \\
\end{cases}
\end{equation*}

It can be assumed that $X_{ij}^{(\lambda)} \sim \mathcal{N}(\mu_i,\sigma_i)$ for $i=1,2,3$ ($j=1,\ldots,n_i$), which represents the distributions of each class.

The maximum likelihood estimate (MLE) of the transformation parameter $\lambda$ , denoted by $\widehat{\lambda}$, is obtained from the maximization of the profile log-likelihood function given by

\begin{equation*}
\small
\begin{aligned}
l(\lambda) = & -\frac{n_1}{2} \log \left( \frac{\sum_{i=1}^{n_1} \left( X_{1i}^{(\lambda)} - \frac{\sum_{i=1}^{n_1} X_{1i}^{(\lambda)}}{n_1} \right)^2}{n_1} \right) 
-\frac{n_2}{2} \log \left( \frac{\sum_{j=1}^{n_2} \left( X_{2j}^{(\lambda)} - \frac{\sum_{j=1}^{n_2} X_{2j}^{(\lambda)}}{n_2} \right)^2}{n_2} \right) \\
& -\frac{n_3}{2} \log \left( \frac{\sum_{k=1}^{n_3} \left( X_{3k}^{(\lambda)} - \frac{\sum_{k=1}^{n_3} X_{3k}^{(\lambda)}}{n_3} \right)^2}{n_3} \right) 
+ (\lambda - 1) \left(\sum_{i=1}^{n_1} \log(X_{1i}) + \sum_{j=1}^{n_2} \log(X_{2j}) + \sum_{k=1}^{n_3} \log(X_{3k})\right) + c.
\end{aligned}
\end{equation*}
where $c$ is a constant. We denote the estimators obtained after applying the Box-Cox transformation to the data as $\widehat{OVL}_N^{BC}$ and $\widehat{VUS}_N^{BC}$.

In practice, it is not always possible to achieve normality for some data sets even after applying the Box-Cox transformation. In such cases, some authors propose a kernel-based approach based on the Gaussian kernel density. The distribution functions in (\ref{VUS}) and (\ref{OVL}) are replaced by appropriate kernel distribution estimators and the integrals are approximated numerically.

\subsubsection{Kernel estimation}
One way to estimate the OVL and the VUS is using the Gaussian kernel approach. Therefore, the estimation of the probability density function is given by
$$\widehat{f_k}(x)=\frac{1}{n_kh_k} \sum_{i=1}^{n_k} \phi \left (\frac{x-X_{ki}}{h_k} \right)$$
and the cumulative distribution function by
$$\widehat{F_k}(x)=\frac{1}{n_k } \sum_{i=1}^{n_k} \Phi \left (\frac{x-X_{ki}}{h_k} \right),$$
for each class $k=1,2,3$, where $n_k$ represents the number of observations in class $k$, $\Phi(\cdot)$ is the standard Normal distribution function, $\phi(\cdot)$ is its density function, and $h_k$ is the bandwidth in each class.

We consider the asymptotic bandwidth proposed by \hyperref[Silverman]{Silverman (1986)} and used by \hyperref[Kang-Tian]{Kang and Tian (2013)} which is given by

\begin{equation}
h_{k}= \left (\frac{4}{3} \right)^{1/5} n_{k}^{-1/5} \min \left(s_k,\frac{IQR_k}{1.349} \right), \quad k=1,2,3,
\label{parámetro de suavizado}
\end{equation}
where $s_k$ and $IQR_k$ denote the standard deviation and the interquartile range of the $k$th sample, respectively. We denote these estimators by $\widehat{VUS}_K$ and $\widehat{OVL}_K$.


\subsubsection{Non-parametric estimation}
A non-parametric alternative is to consider the non-parametric estimation of the ROC surface, proposed by \hyperref[Li y Zhou (2009)]{Li and Zhou (2009)}, which involves replacing the distribution functions in (\ref{ROC def}) with the empirical distribution functions associated with the samples:
\begin{equation}
\widehat{ROC_s}(p_1,p_3)=
\begin{cases}
  \widehat{F_2}\left(\widehat{F_{3}}^{-1}(1-p_3)\right)-\widehat{F_2}\left(\widehat{F_{1}}^{-1}(p_1)\right) & \text{if} \quad \widehat{F_{1}}^{-1}(p_1) \leq \widehat{F_{3}}^{-1}(1-p_3),  \\
  0 & \text{otherwise}. \\
\end{cases}
\end{equation} 
Then:
$$\widehat{VUS}_E=\int_{0}^{1} \int_{0}^{1-\widehat{F_3}\left(\widehat{F_{1}}^{-1}(p_1)\right)} \widehat{{ROC}_s}(p_1,p_3) \, dp_3 \ dp_1.$$

\section{Application to a real data} 

Alzheimer's is the most common degenerative dementia, affecting approximately 47\% of the population aged 85 or older. Neuropsychometric tests have been used to detect Mild Cognitive Impairment (MCI) compared to normal aging and Alzheimer's. The database we have considered includes individuals from a longitudinal cohort of the Alzheimer's Disease Research Center at the University of Washington (ADRC) (\hyperref[Morris (1993)]{Morris, 1993}) and  is available in the GitHub repository (\hyperref[GitHub]{Luo and Xiong, 2014}). The severity of each individual's disease was classified according to the Clinical Dementia Rating (CDR). The dataset contains 14 neuropsychometric markers: global factor (FACTOR1), temporal (ktemp), parietal (kpar), frontal (kfront), logical memory (zpsy004), forward digit span (zpsy005), backward digit span (zpsy006), information (zinfo), two measures of visual retention (zbentc, zbentd), Boston naming test (zboston), mental control (zmentcon), word fluency (zworflu), and associative learning (zassc). These markers were measured in 118 individuals aged 75, divided into three diagnostic categories: the non-demented or healthy group ($D^{-}$, N = 45) with CDR = 0; the MCI group ($D^{0}$, N = 44) with CDR = 0.5; and the Alzheimer's group ($D^{+}$, N = 29) with CDR = 1.


Table \ref{Tabla 3.1} shows the means of each of the 14 biomarkers considered for each of the three diagnostic categories.

\renewcommand{\arraystretch}{1.5}
\captionsetup{labelfont=bf}

\begin{table}[H]
\centering
\caption{Means of the biomarkers}
\small
\begin{tabular}{|c|c|c|c|c|c|c|c|}
\hline
&  \textbf{FACTOR1} &  \textbf{ktemp} &  \textbf{kpar} &  \textbf{kfront} &  \textbf{zpsy004} &  \textbf{zpsy005} &  \textbf{zpsy006} \\ \hline
$\mathbf{D^-}$ &  0.57 &  4.08 &  1.80 &  2.87 &  0.73 &  0.58 &  0.55 \\
$\mathbf{D^0}$ &  -1.62 &  -0.99 &  -0.24 &  0.37 &  -0.86 &  -0.21 &  -0.40 \\
$\mathbf{D^+}$ &  -4.20 &  -5.86 &  -2.38 &  -2.68 &  -1.77 &  -1.21 &  -1.82 \\ \hline
\end{tabular}
\label{Tabla 3.1}
\end{table}

\begin{table}[H]
\centering
\caption*{}
\small
\begin{tabular}{|c|c|c|c|c|c|c|c|}
\hline
& \textbf{zinfo} & \textbf{zbentc} & \textbf{zbentd} &  \textbf{zboston} & \textbf{zmentcon} & \textbf{zworflu} & \textbf{zassc} \\ \hline
$\mathbf{D^-}$ & 0.63 & 0.64 & 0.20 & 0.59 & 0.46 & 0.73 & 0.74\\
$\mathbf{D^0}$ & -0.61 & -0.82 & -0.55 & -0.50 & -0.37 & -0.25 & -0.58\\
$\mathbf{D^+}$ & -2.30 & -1.66 & -1.77 & -3.07 & -1.72 & -1.44 & -1.50\\ \hline
\end{tabular}
\label{Tabla 3.2}
\end{table}

Observe that the mean values per group for each of the 14 biomarkers are ordered monotonically in decreasing order from $D^-$ to $D^+$. Since it's important for the correct calculation of the VUS that the data follows a monotonically increasing order, we assign the opposite sign to the biomarker values.

\begin{figure}[H]
    \centering
    \includegraphics[width=1\linewidth]{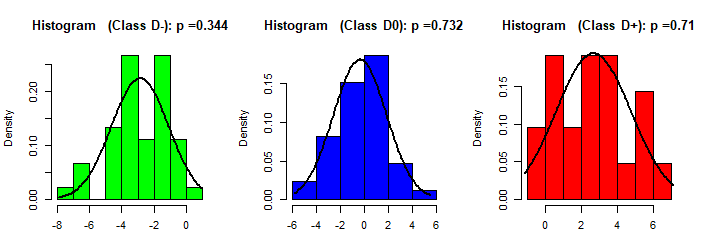}
    \caption{Histograms and p-values of the Shapiro-Wilk's Normality test by diagnostic group for the biomarker \textit{kfront}}
    \label{histogramas kfront}
\end{figure}

Following the proposal by \hyperref[DiagTest]{Luo and Xiong (2012)}, we will select the biomarker \textit{kfront} for our analysis. First, we examine its normality by drawing the corresponding histograms and calculating the p-values of the Shapiro-Wilk's test (see Figure \ref{histogramas kfront}). According to this, we cannot reject the null hypothesis of normality for the biomarker \textit{kfront} for any of the three diagnostic groups $D^-$, $D^0$, and $D^+$. Therefore, we can assume normality when calculating $\widehat{OVL}$ and $\widehat{VUS}$.

The estimators of the VUS and OVL introduced in Section \ref{Section 2} are shown in Table \ref{OVL y VUS kfront}. We have also included a 95-bootstrap percentile confidence interval (95-CI).

\begin{table}[H]
\centering
\caption{Estimations of OVL and VUS for the biomarker \textit{kfront}}
\label{OVL y VUS kfront}
\small
    \begin{tabular}{|c|c|c|c|c|c|} \hline 
         & $\mathbf{\widehat{OVL}_N}$ & $\mathbf{\widehat{VUS}_N}$ & $\mathbf{\widehat{OVL}_K}$ & $\mathbf{\widehat{VUS}_K}$ & $\mathbf{\widehat{VUS}_E}$ \\ \hline  
         \textbf{Point estimator}& 0.1483 & 0.6568 & 0.1870 & 0.6166 & 0.6036 \\ \hline 
         \textbf{95-CI}& (0.0650,0.2279) & (0.5611,0.7585) & (0.0795,0.2783) & (0.5086,0.7238) & (0.4549,0.7407) \\ \hline 
    \end{tabular}
\end{table}

\begin{figure}[H]
    \centering
    \includegraphics[width=1\linewidth]{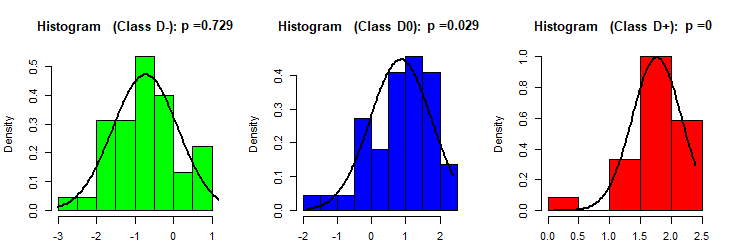}
    \caption{Histograms and p-values of the Shapiro-Wilk's Normality test by diagnostic group for the biomarker \textit{zpsy004}}
    \label{histogramas zpsy004}
\end{figure}

Now,for illustration purposes we have also considered the biomarker \textit{zpsy004}. In Figure \ref{histogramas zpsy004}, we present the histograms and the Shapiro-Wilk's normality test p-values. This analysis lead us to not reject the hypothesis of normality for the group of healthy individuals, $D^-$, and to reject it for those with Mild Cognitive Impairment, belonging to the $D^0$ group, and those with Alzheimer's, members of $D^+$. In this case, as we cannot assume normality for two out of the three diagnostic groups, we obtain the parametric estimations of the measures after Box-Cox transformation. The results can be found in Table \ref{OVL y VUS zpsy004}.

\begin{table}[H]
    \centering
\caption{Estimations of OVL and VUS for the biomarker \textit{zpsy004}}
\label{OVL y VUS zpsy004}
\small
    \begin{tabular}{|c|c|c|c|c|c|} \hline 
         & $\mathbf{\widehat{OVL}_N^{BC}}$ & $\mathbf{\widehat{VUS}_N^{BC}} $ & $\mathbf{\widehat{OVL}_K}$ & $\mathbf{\widehat{VUS}_K}$ & $\mathbf{\widehat{VUS}_E}$ \\ \hline
         \textbf{Point estimator} & 0.0424 & 0.7242 & 0.0889 & 0.6715 & 0.7628 \\ \hline 
         \textbf{95-PB CI} & (0.0054,0.1077) & (0.6372,0.8153) & (0.0098,0.1792) & (0.5778,0.7801) & (0.6494,0.8576) \\ \hline 
    \end{tabular}
\end{table}

The results indicate that \textit{zpsy004} shows a greater discriminatory capacity compared to \textit{kfront}, both in terms of point estimates and the 95\% percentile bootstrap confidence intervals obtained.

\section{Simulation study}
The aim of this section is two-fold. Firstly, we evaluate the bias, the Root Mean Squared Error (RMSE) and the coverages of 95\% percentile confidence intervals of the estimators proposed on Section~2. Secondly, we compare the performance of OVL with VUS in order to assess analogous testing procedures. R software was employed throughout (https://www.r-project.org/, R Core Team,
Vienna, Austria). R code for the estimation of OVL is provided under request.

\subsection{Assessment of the accuracy and bias of the estimators of the
OVL}

A Monte Carlo simulation study of size 1000 was performed in order to evaluate the finite sample properties of the
two estimators of OVL, namely $\widehat{OVL}_N$ and $\widehat{OVL}_K$ in terms of bias, RMSE and 95\%-coverages. We consider several
distributional scenarios. These are specified in Table~\ref{TT1}. Sample sizes considered were $n_1 =n_2=n_3 = 20$, $50$, and $100$.

Results are shown in Table~\ref{TT2}. In general, the estimator $\widehat{OVL}_K$ tends to provide better performance in terms of precision (RMSE) and 95\%-coverages. This trend holds across most scenarios, with the exception of scenarios 5 and 6, where $\widehat{OVL}_N$ performs comparably or slightly better.  Regarding bias, both estimators exhibit a consistent tendency toward minimizing bias as the sample size increases, with bias values approaching zero for larger sample sizes (e.g., at $n=100$). In general, both methods provide robust results.

\begin{table}[tp]
\caption{Distributions used in the simulation study of Section 4.1.}
\centering
\footnotesize
\begin{tabular}{l}
\toprule
\textbf{Scenarios} \\ 
\hline
1. $F_1 \sim \mathcal{N}(0,1)$, $F_2 \sim \mathcal{N}(1/2,1)$ and $F_3 \sim \mathcal{N}(1,1)$ \\ 
2. $F_1 \sim \mathrm{LogN}(0,1)$, $F_2 \sim \mathrm{LogN}(0,1)$ and $F_3 \sim \mathrm{LogN}(1,1)$ \\ 
3. $F_1 \sim \Gamma(2,1)$, $F_2 \sim \Gamma(3,1)$ and $F_3 \sim \Gamma(4,1)$ \\ 
4. $F_1 \sim N(0,1)$, $F_2 \sim \Gamma(2,1)$ and $F_3 \sim \mathrm{LogN}(0,1)$ \\ 
5. $F_1 \sim \frac{1}{2}\mathcal{N}(0,1) + \frac{1}{2}\mathcal{N}(3,1)$, $F_2 \sim \frac{1}{2}\mathcal{N}(1,1) + \frac{1}{2}\mathcal{N}(4,3/2)$ and $F_3 \sim \frac{1}{2}\mathcal{N}(2,1) + \frac{1}{2}\mathcal{N}(5,2)$ \\ 
6. $F_1 \sim \frac{1}{2}\Gamma(1,1) + \frac{1}{2}\Gamma(4,1)$, $F_2 \sim \frac{1}{2}\Gamma(2,1) + \frac{1}{2}\Gamma(5,2/3)$ and $F_3 \sim \frac{1}{2}\Gamma(3,1) + \frac{1}{2}\Gamma(6,1/2)$ \\ 
7. $F_1 \sim \frac{1}{2}\mathcal{N}(0,1) + \frac{1}{2}\Gamma(4,1)$, $F_2 \sim \frac{1}{2}\mathcal{N}(1,1) + \frac{1}{2}\Gamma(5,2/3)$ and $F_3 \sim \frac{1}{2}\mathcal{N}(2,1) + \frac{1}{2}\Gamma(6,1/2)$ \\ 
\bottomrule
\hline
\end{tabular}\label{TT1}
\end{table}

\begin{table}[tp]
\caption{Bias, and RMSE and 95\%-coverages for the scenarios of Table~\ref{TT1}}
\centering
\footnotesize
\begin{tabular}{clccccc ccc ccc}
\toprule
 & & & \multicolumn{3}{c}{\textbf{Bias}} & \multicolumn{3}{c}{\textbf{RMSE}} & \multicolumn{3}{c}{\textbf{95\%-Coverage}} \\
\cline{4-12}
Scenarios & & {\small Sample size} & 20 & 50 & 100 & 20 & 50 & 100 & 20 & 50 & 100 \\
\hline
1. & $\widehat{OVL}_N$ &  & \textbf{-0.029} & \textbf{-0.005} & \textbf{-0.002} & 0.107 & 0.072 & 0.053 & \textbf{0.871} & 0.930 & 0.939 \\
 & $\widehat{OVL}_K$ &  & -0.031 & 0.008 & 0.015 & \textbf{0.099} & \textbf{0.068} & \textbf{0.055} & 0.842 & \textbf{0.944} & \textbf{0.955 }\\
\hline
2. & $\widehat{OVL}_N$ &  & -0.076 & -0.039 & -0.027 & 0.129 & 0.076 & 0.054 & \textbf{0.738} & 0.808 & 0.857 \\
 & $\widehat{OVL}_K$ &  & \textbf{-0.065} & \textbf{-0.025} & \textbf{-0.015} & \textbf{0.116} & \textbf{0.070} & \textbf{0.051} & 0.737 & \textbf{0.848} & \textbf{0.883} \\
\hline
3. & $\widehat{OVL}_N$ &  & -0.026 & \textbf{-0.016} & \textbf{-0.008} & 0.111 & \textbf{0.074} & \textbf{0.049} & 0.883 & 0.915 & 0.931 \\
 & $\widehat{OVL}_K$ &  & \textbf{-0.003} & \textbf{0.016} & 0.022 & \textbf{0.100} & \textbf{0.074} & 0.054 & \textbf{0.919} & \textbf{0.965} & \textbf{0.950}\\
\hline
4. & $\widehat{OVL}_N$ &  & -0.037 & \textbf{-0.021} & \textbf{-0.014} & \textbf{0.083} &\textbf{ 0.051} & \textbf{0.037} & 0.857 & 0.864 & 0.878 \\
 & $\widehat{OVL}_K$ &  & \textbf{0.018} & 0.029 & 0.028 & 0.088 & 0.062 & 0.050 & \textbf{0.883} & \textbf{0.944} & \textbf{0.920} \\
\hline
5. & $\widehat{OVL}_N$ &  & \textbf{0.007} & \textbf{0.022} & \textbf{0.028} & 0.102 & \textbf{0.070} & \textbf{0.055} & 0.935 & \textbf{0.945} & \textbf{0.899} \\
 & $\widehat{OVL}_K$ &  & 0.039 & 0.072 & 0.078 & \textbf{0.094} & 0.091 & 0.088 & \textbf{0.966} & 0.934 & 0.688 \\
\hline
6. & $\widehat{OVL}_N$ &  & \textbf{-0.043} & \textbf{-0.010} & \textbf{-0.000} & \textbf{0.110} &\textbf{ 0.067} &\textbf{ 0.048} & \textbf{0.820} & \textbf{0.901} & \textbf{0.918} \\
 & $\widehat{OVL}_K$ &  & -0.072 & -0.028 & -0.017 & 0.118 & 0.070 & 0.053 & 0.665 & 0.801 & 0.883 \\
\hline
7. & $\widehat{OVL}_N$ &  & \textbf{0.009} & 0.037 & 0.049 & \textbf{0.093} & 0.072 & 0.066 & \textbf{0.908} & 0.913 & 0.813 \\
 & $\widehat{OVL}_K$ &  & -0.033 & \textbf{-0.001} & \textbf{0.008} & 0.095 & \textbf{0.060} & \textbf{0.044} & 0.828 & \textbf{0.916} & \textbf{0.938} \\
\hline
\bottomrule
\end{tabular}
\label{TT2}
\end{table}

\subsection{Comparison of the discriminatory capacity of OVL and VUS estimators}

In this section, we conduct a Monte Carlo simulation study of size 1,000 using different distributions to compare the performance in terms of power of $\widehat{OVL}$ and $\widehat{VUS}$ as methods for evaluating the discriminatory capacity of biomarkers when working with 3 populations. In each simulation, 500 Bootstrap resamples are generated and the significance level is $\alpha=0.05$.

The null hypothesis for the tests based on the two statistics is the same, that the biomarker is not informative (that is, OVL=1 and VUS= 1/6, respectively).

We consider a total of 17 scenarios based on the different distributions $F_1$, $F_2$, and $F_3$. In the scenarios in which $F_1$, $F_2$, and $F_3$ are the same, the rejection rate corresponds to the Type I error, and in the rest, to the power of $\widehat{OVL}$ and $\widehat{VUS}$.

Within each scenario, 8 different sample sizes have been considered: $n_x=n_y=n_z=20, 50, 100$ and $(n_x,n_y,n_z)=(20, 20, 30), (20, 30, 50), (30, 50, 50), (50, 50, 100)$ and $(50, 100, 100)$

\subsection{Normal distributions}\label{Normal}
We begin the simulation study assuming that the three classes follow normal distributions.

First, we assume $F_1 \sim \mathcal{N}(0,1)$, $F_2 \sim \mathcal{N}(0,1)$, and $F_3 \sim \mathcal{N}(0,1)$; we are actually under the null hypothesis $H_0: F_1=F_2=F_3$.
The theoretical OVL reaches the maximum value, unity. The theoretical value of the VUS, on the other hand, is 1/6, which is the minimum value it can attain and corresponds to the volume of the tetrahedron.

In Table \ref{Tabla 1}, the rejection proportions for the case under study are presented. Since the three classes are identically distributed, so we expect the calculated values to be around $\alpha=0.05$. Generally, except a few cases that are conservative, the rejection proportions of both statistics are close to the nominal.

\begin{table}[H]
\centering
\renewcommand{\arraystretch}{1.25}
\caption{Proportions of rejection for $F_1 \sim \mathcal{N}(0,1)$, $F_2 \sim \mathcal{N}(0,1)$ and $F_3 \sim \mathcal{N}(0,1)$}
\small
\begin{tabular}{|l|l|l|l|l|l|}
\hline
\multicolumn{1}{|c|}{\textbf{$\mathbf{(n_x,n_y,n_z)}$}} & $\mathbf{\widehat{OVL}_N}$ & $\mathbf{\widehat{VUS}_N}$ & $\mathbf{\widehat{OVL}_K}$ & $\mathbf{\widehat{VUS}_K}$ & $\mathbf{\widehat{VUS}_E} $ \\ \hline
$\mathbf{(20, 20, 20)}$ & 0.058 & 0.055 & 0.045 & 0.055 & 0.044 \\
$\mathbf{(20, 20, 30)}$ & 0.056 & 0.046 & 0.042 & 0.052 & 0.055 \\
$\mathbf{(20, 30, 50)}$ & 0.037 & 0.049 & 0.028 & 0.053 & 0.049\\
$\mathbf{(30, 50, 50)}$ & 0.049 & 0.043 & 0.045 & 0.047 & 0.049 \\
$\mathbf{(50, 50, 50)}$ & 0.052 & 0.036 & 0.045 & 0.037 & 0.036 \\ 
$\mathbf{(50, 50, 100)}$ & 0.045 & 0.057 & 0.032 & 0.053 & 0.057 \\ 
$\mathbf{(50, 100, 100)}$ & 0.058 & 0.054 & 0.049 & 0.052 & 0.053 \\
$\mathbf{(100, 100, 100)}$ & 0.049 & 0.046 & 0.037 & 0.050 & 0.050 \\ \hline
\end{tabular}
\label{Tabla 1}
\end{table}

In the second scenario, $F_1 \sim \mathcal{N}(0,1)$, $F_2 \sim \mathcal{N}(1/2,1)$, and $F_3 \sim \mathcal{N}(1,1)$.
The theoretical OVL is 0.6171, while the theoretical VUS is 0.3372.

Table \ref{Tabla 2} shows that the test power is higher for $\widehat{VUS}$ than for $\widehat{OVL}$ regardless of the estimation method used, with the difference being most evident when the sample size is smaller and when kernel estimation is used: $\widehat{VUS}$ rejects almost 90\% of the time for the smallest sample sizes considered ($n_x=20, n_y=20, n_z=20$), while $\widehat{OVL}$ does so in less than 50\% of these cases. As expected, parametric estimations are the best. However, the results for $\widehat{OVL}$ improve considerably for larger sample sizes, with both measures achieving maximum power when all three classes reach a sample size of 100.

\begin{table}[H]
\centering
\renewcommand{\arraystretch}{1.25}
\caption{Proportions of rejection for $F_1 \sim \mathcal{N}(0,1)$, $F_2 \sim \mathcal{N}(1/2,1)$ and $F_3 \sim \mathcal{N}(1,1)$}
\small
\begin{tabular}{|l|l|l|l|l|l|}
\hline
\multicolumn{1}{|c|}{\textbf{$\mathbf{(n_x,n_y,n_z)}$}} & $\mathbf{\widehat{OVL}_N}$ & $\mathbf{\widehat{VUS}_N}$ & $\mathbf{\widehat{OVL}_K}$ & $\mathbf{\widehat{VUS}_K}$ & $\mathbf{\widehat{VUS}_E} $ \\ \hline
$\mathbf{(20, 20, 20)}$ & 0.724 & \textbf{0.890} & 0.472 & 0.874 & 0.861 \\
$\mathbf{(20, 20, 30)}$ & 0.779 & \textbf{0.922} & 0.504 & 0.901 & 0.895 \\
$\mathbf{(20, 30, 50)}$ & 0.872 & \textbf{0.973} & 0.644 & 0.960 & 0.953 \\
$\mathbf{(30, 50, 50)}$ & 0.965 & \textbf{0.991} & 0.840 & 0.988 & 0.986 \\
$\mathbf{(50, 50, 50)}$ & 0.995 & \textbf{0.999} & 0.937 & 0.997 & 0.996 \\
$\mathbf{(50, 50, 100)}$ & 0.997 & \textbf{1.000} & 0.978 & \textbf{1.000} & \textbf{1.000} \\ 
$\mathbf{(50, 100, 100)}$ & 0.998 & \textbf{1.000} & 0.987 & \textbf{1.000} & 0.999 \\
$\mathbf{(100, 100, 100)}$ & \textbf{1.000} & \textbf{1.000} & \textbf{1.000} & \textbf{1.000} & \textbf{1.000} \\ \hline
\end{tabular}
\label{Tabla 2}
\end{table}

In the third scenario, we assume that the normal distributions differ not in mean but in standard deviation: $F_1 \sim \mathcal{N}(0,1)$, $F_2 \sim \mathcal{N}(0,3/2)$, and $F_3 \sim \mathcal{N}(0,2)$.
The theoretical values of OVL and VUS in this new scenario are 0.6773 and 0.1668, respectively.

Table \ref{Tabla 3} shows results that are completely different from those in Table \ref{Tabla 2}. $\widehat{OVL}$ achieves a much higher power than $\widehat{VUS}$, and its rejection rate improves significantly as the sample size increases. $\widehat{VUS}$, however, faces serious difficulties in rejecting the null hypothesis, a problem that does not improve with increasing sample size.

\begin{table}[H]
\centering
\renewcommand{\arraystretch}{1.25}
\caption{Proportions of rejection for $F_1 \sim \mathcal{N}(0,1)$, $F_2 \sim \mathcal{N}(0,3/2)$ and $F_3 \sim \mathcal{N}(0,2)$}
\small
\begin{tabular}{|l|l|l|l|l|l|}
\hline
\multicolumn{1}{|c|}{\textbf{$\mathbf{(n_x,n_y,n_z)}$}} & $\mathbf{\widehat{OVL}_N}$ & $\mathbf{\widehat{VUS}_N}$ & $\mathbf{\widehat{OVL}_K}$ & $\mathbf{\widehat{VUS}_K}$ & $\mathbf{\widehat{VUS}_E} $ \\ \hline
$\mathbf{(20, 20, 20)}$ & \textbf{0.469} & 0.058 & 0.315 & 0.053 & 0.057 \\
$\mathbf{(20, 20, 30)}$ & \textbf{0.547} & 0.040 & 0.375 & 0.051 & 0.050 \\
$\mathbf{(20, 30, 50)}$ & \textbf{0.727} & 0.040 & 0.512 & 0.039 & 0.045 \\
$\mathbf{(30, 50, 50)}$ & \textbf{0.902} & 0.031 & 0.727 & 0.035 & 0.036 \\
$\mathbf{(50, 50, 50)}$ & \textbf{0.959} & 0.034 & 0.844 & 0.034 & 0.039 \\ 
$\mathbf{(50, 50, 100)}$ & \textbf{0.994} & 0.042 & 0.939 & 0.048 & 0.047 \\ 
$\mathbf{(50, 100, 100)}$ & \textbf{0.996} & 0.047 & 0.969 & 0.045 & 0.046 \\
$\mathbf{(100, 100, 100)}$ & \textbf{0.999} & 0.041 & 0.996 & 0.045 & 0.048 \\ \hline
\end{tabular}
\label{Tabla 3}
\end{table}

\subsection{Log-Normal distributions}
The Log-Normal distribution is positively skewed and is characterized by the fact that its natural logarithm follows a Normal distribution. Mathematically, if $X \sim \mathrm{LogN}(\mu,\sigma)$, then $\log(X) \sim \mathcal{N}(\mu,\sigma)$. Within this subsection, we will consider three additional scenarios to evaluate the discriminatory power of the two metrics we are comparing, following the structure of Section \ref{Normal}: a first scenario under $H_0$ and two under the alternative, modifying the means and standard deviations of the Log-Normal distribution.

Let's consider that the three classes are identically distributed as $\mathrm{LogN}(0,1)$, that is, $F_1 \sim \mathrm{LogN}(0,1)$, $F_2 \sim \mathrm{LogN}(0,1)$, and $F_3 \sim \mathrm{LogN}(0,1)$. We observe that the theoretical results of the first scenario are repeated, where $OVL=1$ and $VUS=1/6$.

Table \ref{Tabla 4} shows that both $\widehat{VUS}$ and $\widehat{OVL}$ achieve values close to the significance level $\alpha$ for all sample sizes and estimation methods analyzed. $\widehat{OVL}_K$ is sometimes conservative.

\begin{table}[H]
\centering
\renewcommand{\arraystretch}{1.25}
\caption{Proportions of rejection for $F_1 \sim \mathrm{LogN}(0,1)$, $F_2 \sim \mathrm{LogN}(0,1)$ and $F_3 \sim \mathrm{LogN}(0,1)$}
\small
\begin{tabular}{|l|l|l|l|l|l|}
\hline
\multicolumn{1}{|c|}{\textbf{$\mathbf{(n_x,n_y,n_z)}$}} & $\mathbf{\widehat{OVL}_N^{BC}}$ & $\mathbf{\widehat{VUS}_N^{BC}}$ & $\mathbf{\widehat{OVL}_K}$ & $\mathbf{\widehat{VUS}_K}$ & $\mathbf{\widehat{VUS}_E} $ \\ \hline
$\mathbf{(20, 20, 20)}$ & 0.070 & 0.054 & 0.032 & 0.058 & 0.044 \\
$\mathbf{(20, 20, 30)}$ & 0.053 & 0.045 & 0.039 & 0.050 & 0.055 \\
$\mathbf{(20, 30, 50)}$ & 0.040 & 0.049 & 0.028 & 0.052 & 0.049 \\
$\mathbf{(30, 50, 50)}$ & 0.052 & 0.040 & 0.045 & 0.043 & 0.049 \\
$\mathbf{(50, 50, 50)}$ & 0.047 & 0.035 & 0.035 & 0.043 & 0.036 \\ 
$\mathbf{(50, 50, 100)}$ & 0.044 & 0.054 & 0.036 & 0.054 & 0.057 \\ 
$\mathbf{(50, 100, 100)}$ & 0.061 & 0.053 & 0.047 & 0.051 & 0.053 \\
$\mathbf{(100, 100, 100)}$ & 0.052 & 0.044 & 0.045 & 0.054 & 0.050 \\ \hline
\end{tabular}
\label{Tabla 4}
\end{table}

In the following scenario, we slightly modify the previous distributions, altering only the mean of the distribution of the third class: $F_1 \sim \mathrm{LogN}(0,1)$, $F_2 \sim \mathrm{LogN}(0,1)$ and $F_3 \sim \mathrm{LogN}(1,1)$.
For this scenario, $OVL=0.6171$ and $VUS=0.3169$.

The results in Table \ref{Tabla 5} show that $\widehat{OVL}_N^{BC}$ is the best and only overperformed by $\widehat{OVL}_K$, with a slight better performance for the small and equal sample sizes.

Likewise, it is worth highlighting that, despite only one of the distributions being different,  which makes the discrimination by statistical tests more challenging, both $\widehat{OVL}$ and $\widehat{VUS}$ adequately detect the difference between distributions and reject, in the worst case, in more than 60 out of 100 simulations performed.

\begin{table}[H]
\centering
\renewcommand{\arraystretch}{1.25}
\caption{Proportions of rejection for $F_1 \sim \mathrm{LogN}(0,1)$, $F_2 \sim \mathrm{LogN}(0,1)$ and $F_3 \sim \mathrm{LogN}(1,1)$}
\small
\begin{tabular}{|l|l|l|l|l|l|}
\hline
\multicolumn{1}{|c|}{\textbf{$\mathbf{(n_x,n_y,n_z)}$}} & $\mathbf{\widehat{OVL}_N^{BC}}$ & $\mathbf{\widehat{VUS}_N^{BC}}$ & $\mathbf{\widehat{OVL}_K}$ & $\mathbf{\widehat{VUS}_K}$ & $\mathbf{\widehat{VUS}_E} $ \\ \hline
$\mathbf{(20, 20, 20)}$ & 0.837 & 0.836 & 0.612 & \textbf{0.846} & 0.795 \\
$\mathbf{(20, 20, 30)}$ & \textbf{0.901} & 0.856 & 0.650 & 0.886 & 0.811 \\
$\mathbf{(20, 30, 50)}$ & \textbf{0.975} & 0.922 & 0.766 & 0.956 & 0.893 \\
$\mathbf{(30, 50, 50)}$ & \textbf{0.994} & 0.964 & 0.944 & 0.985 & 0.958 \\
$\mathbf{(50, 50, 50)}$ & \textbf{1.000} & 0.995 & 0.981 & 0.996 & 0.987 \\
$\mathbf{(50, 50, 100)}$ & \textbf{1.000} & 0.998 & 0.996 & \textbf{1.000} & 0.994 \\ 
$\mathbf{(50, 100, 100)}$ & \textbf{1.000} & 0.997 & 0.999 & 0.999 & 0.997 \\
$\mathbf{(100, 100, 100)}$ & \textbf{1.000} & \textbf{1.000} & \textbf{1.000} & \textbf{1.000} & \textbf{1.000} \\ \hline
\end{tabular}
\label{Tabla 5}
\end{table}

We propose, in the next scenario, a situation in which the three distributions have the same mean but different standard deviations: $F_1 \sim \mathrm{LogN}(1,1/2)$, $F_2 \sim \mathrm{LogN}(1,1)$ and $F_3 \sim \mathrm{LogN}(1,3/2)$. The theoretical overlap measure gives us $OVL=0.5157$, while the theoretical volume under the ROC surface is $VUS=0.1674$.

As it was the case with Table \ref{Tabla 3} for the Normal distribution when the distributions differed in standard deviation, Table \ref{Tabla 6} for this scenario shows that $\widehat{OVL}$ has a much higher discriminative power than $\widehat{VUS}$ for all sample sizes considered in this work.

Moreover, the power associated with $\widehat{OVL}_N^{BC}$ rises as the sample size increases, reaching the maximum possible power (unity) for the moderate and large sample sizes analyzed ($n_x=100, n_y=100, n_z=100$), whereas $\widehat{VUS}$ does not exceed 0.276 in the best case.

\begin{table}[H]
\centering
\renewcommand{\arraystretch}{1.25}
\caption{Proportions of rejection for $F_1 \sim \mathrm{LogN}(1,1/2)$, $F_2 \sim \mathrm{LogN}(1,1)$ and $F_3 \sim \mathrm{LogN}(1,3/2)$}
\small
\begin{tabular}{|l|l|l|l|l|l|}
\hline
\multicolumn{1}{|c|}{\textbf{$\mathbf{(n_x,n_y,n_z)}$}} & $\mathbf{\widehat{OVL}_N^{BC}}$ & $\mathbf{\widehat{VUS}_N^{BC}}$ & $\mathbf{\widehat{OVL}_K}$ & $\mathbf{\widehat{VUS}_K}$ & $\mathbf{\widehat{VUS}_E} $ \\ \hline
$\mathbf{(20, 20, 20)}$ & \textbf{0.938} & 0.057 & 0.401 & 0.188 & 0.069 \\
$\mathbf{(20, 20, 30)}$ & \textbf{0.978} & 0.043 & 0.334 & 0.198 & 0.051 \\
$\mathbf{(20, 30, 50)}$ & \textbf{0.992} & 0.037 & 0.332 & 0.200 & 0.043 \\
$\mathbf{(30, 50, 50)}$ & \textbf{1.000} & 0.035 & 0.728 & 0.225 & 0.036 \\
$\mathbf{(50, 50, 50)}$ & \textbf{1.000} & 0.035 & 0.928 & 0.226 & 0.053 \\
$\mathbf{(50, 50, 100)}$ & \textbf{1.000} & 0.042 & 0.932 & 0.275 & 0.050 \\ 
$\mathbf{(50, 100, 100)}$ & \textbf{1.000} & 0.040 & 0.979 & 0.275 & 0.046 \\
$\mathbf{(100, 100, 100)}$ & \textbf{1.000} & 0.048 & \textbf{1.000} & 0.276 & 0.054 \\ \hline
\end{tabular}
\label{Tabla 6}
\end{table}

\subsection{Gamma distributions}

The Gamma distribution, like the Log-Normal distribution, is asymmetric and only takes positive values. It has two parameters: the shape, $\alpha > 0$, and the scale, $\beta > 0$. For the analysis of their rejection proportions, we consider three scenarios.

First, we will analyze the behavior of $\widehat{OVL}$ and $\widehat{VUS}$ when all three classes follow the same distribution: $\Gamma(\alpha=1,\beta=1)$. The theoretical values for the OVL and VUS are 1 and 1/6, respectively.

Table \ref{Tabla 7} presents the rejection proportions for this scenario. Since we are under the null hypothesis $H_0: F_1 = F_2 = F_3$, as expected, the observed values are more or less close to the significance level $\alpha = 0.05$.

\begin{table}[H]
\centering
\renewcommand{\arraystretch}{1.25}
\caption{Proportions of rejection for $F_1 \sim \Gamma(1,1)$, $F_2 \sim \Gamma(1,1)$ and $F_3 \sim \Gamma(1,1)$}
\small
\begin{tabular}{|l|l|l|l|l|l|}
\hline
\multicolumn{1}{|c|}{\textbf{$\mathbf{(n_x,n_y,n_z)}$}} & $\mathbf{\widehat{OVL}_N^{BC}}$ & $\mathbf{\widehat{VUS}_N^{BC}}$ & $\mathbf{\widehat{OVL}_K}$ & $\mathbf{\widehat{VUS}_K}$ & $\mathbf{\widehat{VUS}_E} $ \\ \hline
$\mathbf{(20, 20, 20)}$ & 0.055 & 0.059 & 0.033 & 0.055 & 0.046 \\
$\mathbf{(20, 20, 30)}$ & 0.063 & 0.069 & 0.039 & 0.063 & 0.064 \\
$\mathbf{(20, 30, 50)}$ & 0.054 & 0.047 & 0.045 & 0.048 & 0.048 \\
$\mathbf{(30, 50, 50)}$ & 0.064 & 0.049 & 0.041 & 0.047 & 0.050 \\
$\mathbf{(50, 50, 50)}$ & 0.045 & 0.047 & 0.027 & 0.048 & 0.046 \\
$\mathbf{(50, 50, 100)}$ & 0.055 & 0.054 & 0.042 & 0.055 & 0.057 \\ 
$\mathbf{(50, 100, 100)}$ & 0.064 & 0.049 & 0.055 & 0.052 & 0.051 \\
$\mathbf{(100, 100, 100)}$ & 0.058 & 0.071 & 0.048 & 0.063 & 0.069 \\ \hline
\end{tabular}
\label{Tabla 7}
\end{table}

In the next scenario, the classes come from distributions with the same scale parameter but different shape parameters: $F_1 \sim \Gamma(2,1)$, $F_2 \sim \Gamma(3,1)$, and $F_3 \sim \Gamma(4,1)$. In this scenario, the values for OVL and VUS are $OVL = 0.5295$ and $VUS = 0.3888$.

As shown in Table \ref{Tabla 8}, $\widehat{VUS}_N^{BC}$ achieves slightly higher powers than $\widehat{OVL}_N^{BC}$ for all sample sizes, with the differences being greater when the kernel method is used.

\begin{table}[H]
\centering
\renewcommand{\arraystretch}{1.25}
\caption{Proportions of rejection for $F_1 \sim \Gamma(2,1)$, $F_2 \sim \Gamma(3,1)$ and $F_3 \sim \Gamma(4,1)$}
\small
\begin{tabular}{|l|l|l|l|l|l|}
\hline
\multicolumn{1}{|c|}{\textbf{$\mathbf{(n_x,n_y,n_z)}$}} & $\mathbf{\widehat{OVL}_N^{BC}}$ & $\mathbf{\widehat{VUS}_N^{BC}}$ & $\mathbf{\widehat{OVL}_K}$ & $\mathbf{\widehat{VUS}_K}$ & $\mathbf{\widehat{VUS}_E} $ \\ \hline
$\mathbf{(20, 20, 20)}$ & 0.910 & \textbf{0.980} & 0.672 & 0.959 & 0.967 \\
$\mathbf{(20, 20, 30)}$ & 0.946 & \textbf{0.990} & 0.749 & 0.978 & 0.978 \\
$\mathbf{(20, 30, 50)}$ & 0.978 & \textbf{0.996} & 0.871 & 0.993 & 0.992 \\
$\mathbf{(30, 50, 50)}$ & 0.999 & \textbf{1.000} & 0.963 & \textbf{1.000} & \textbf{1.000} \\
$\mathbf{(50, 50, 50)}$ & \textbf{1.000} & \textbf{1.000} & 0.996 & \textbf{1.000} & \textbf{1.000} \\
$\mathbf{(50, 50, 100)}$ & \textbf{1.000} & \textbf{1.000} & \textbf{1.000} & \textbf{1.000} & \textbf{1.000} \\ 
$\mathbf{(50, 100, 100)}$ & \textbf{1.000} & \textbf{1.000} & 0.999 & \textbf{1.000} & \textbf{1.000} \\
$\mathbf{(100, 100, 100)}$ & \textbf{1.000} & \textbf{1.000} & \textbf{1.000} & \textbf{1.000} & \textbf{1.000} \\ \hline
\end{tabular}
\label{Tabla 8}
\end{table}

In the last scenario, both the shape and scale parameters differ between populations: $F_1 \sim \Gamma(0.2,0.6)$, $F_2 \sim \Gamma(0.2,0.7)$ and $F_3 \sim \Gamma(0.5,0.5)$. The theoretical values are 0.6138 for OVL and 0.3056 for VUS.

The rejection proportions are presented in Table \ref{Tabla 9}, and we observe that $\widehat{VUS}$ outperforms $\widehat{OVL}$ in terms of power when the kernel estimation is chosen, but the opposite occurs with the parametric method.

\begin{table}[H]
\centering
\renewcommand{\arraystretch}{1.25}
\caption{Proportions of rejection for $F_1 \sim \Gamma(0.2,0.6)$, $F_2 \sim \Gamma(0.2,0.7)$ and $F_3 \sim \Gamma(0.5,0.5)$}
\begin{tabular}{|l|l|l|l|l|l|}
\hline
\multicolumn{1}{|c|}{\textbf{$\mathbf{(n_x,n_y,n_z)}$}} & $\mathbf{\widehat{OVL}_N^{BC}}$ & $\mathbf{\widehat{VUS}_N^{BC}}$ & $\mathbf{\widehat{OVL}_K}$ & $\mathbf{\widehat{VUS}_K}$ & $\mathbf{\widehat{VUS}_E} $ \\ \hline
$\mathbf{(20, 20, 20)}$ & \textbf{0.741} & 0.646 & 0.244 & 0.457 & 0.550 \\
$\mathbf{(20, 20, 30)}$ & \textbf{0.821} & 0.741 & 0.388 & 0.512 & 0.635 \\
$\mathbf{(20, 30, 50)}$ & \textbf{0.896} & 0.786 & 0.527 & 0.597 & 0.685 \\
$\mathbf{(30, 50, 50)}$ & \textbf{0.980} & 0.873 & 0.590 & 0.732 & 0.809 \\
$\mathbf{(50, 50, 50)}$ & \textbf{0.995} & 0.948 & 0.643 & 0.816 & 0.885 \\
$\mathbf{(50, 50, 100)}$ & \textbf{1.000} & 0.964 & 0.839 & 0.905 & 0.922 \\ 
$\mathbf{(50, 100, 100)}$ & \textbf{1.000} & 0.982 & 0.901 & 0.941 & 0.967 \\
$\mathbf{(100, 100, 100)}$ & \textbf{1.000} & \textbf{1.000} & 0.944 & 0.986 & 0.996 \\ \hline
\end{tabular}
\label{Tabla 9}
\end{table}

\subsection{A combination of different distributions}

We now present a simulation in which the three classes are obtained from populations from different distribution families. Specifically, we assume that each of the three classes comes from one of the three distribution families we have considered before. Thus, the random observations of class 1 are normally distributed $F_1 \sim \mathcal{N}(0,1)$, those of class 2 follow a Gamma distribution $F_2 \sim \Gamma(2,1)$, and those of class 3 follow a Log-Normal distribution $F_3 \sim \mathrm{LogN}(0,1)$. For this scenario, the theoretical values are: $OVL=0.3959$ and $VUS=0.2943$.

According to Table \ref{Tabla 10}, the $\widehat{OVL}$ metric achieves higher powers for all the sample sizes considered, obtaining better results with the parametric estimation.
As in previous scenarios, the differences are more significant for small sample sizes. Thus, when working with datasets with few observations, it is most appropriate to use $\widehat{OVL}_N^{BC}$ in this scenario.

\begin{table}[H]
\centering
\renewcommand{\arraystretch}{1.25}
\caption{Proportions of rejection for $F_1 \sim \mathcal{N}(0,1)$, $F_2 \sim \Gamma(2,1)$ and $F_3 \sim \mathrm{LogN}(0,1)$}
\begin{tabular}{|l|l|l|l|l|l|}
\hline
\multicolumn{1}{|c|}{\textbf{$\mathbf{(n_x,n_y,n_z)}$}} & $\mathbf{\widehat{OVL}_N^{BC}}$ & $\mathbf{\widehat{VUS}_N^{BC}} $ & $\mathbf{\widehat{OVL}_K}$ & $\mathbf{\widehat{VUS}_K}$ & $\mathbf{\widehat{VUS}_E} $ \\ \hline
$\mathbf{(20, 20, 20)}$ & \textbf{0.998} & 0.745 & 0.874 & 0.698 & 0.674 \\
$\mathbf{(20, 20, 30)}$ & \textbf{0.999} & 0.779 & 0.892 & 0.751 & 0.714 \\
$\mathbf{(20, 30, 50)}$ & \textbf{1.000} & 0.865 & 0.948 & 0.859 & 0.831 \\
$\mathbf{(30, 50, 50)}$ & \textbf{1.000} & 0.935 & 0.999 & 0.920 & 0.903 \\
$\mathbf{(50, 50, 50)}$ & \textbf{1.000} & 0.973 & \textbf{1.000} & 0.958 & 0.947 \\
$\mathbf{(50, 50, 100)}$ & \textbf{1.000} & 0.992 & \textbf{1.000} & 0.989 & 0.982 \\ 
$\mathbf{(50, 100, 100)}$ & \textbf{1.000} & 0.994 & \textbf{1.000} & 0.989 & 0.989 \\
$\mathbf{(100, 100, 100)}$ & \textbf{1.000} & \textbf{1.000} & \textbf{1.000} & 0.998 & 0.997 \\ \hline
\end{tabular}
\label{Tabla 10}
\end{table}

\subsection{Mixture distributions}

Here we present a total of 7 scenarios: 3 for mixtures of Normals, 2 for mixtures of Gammas, and 2 more for the possibility that the observations come from a mixture of Normals and Gammas. In all of these scenarios, two distributions are combined, each given an equal weight of $1/2$.

First, we will consider that the three classes are identically distributed, with the distribution $\frac{1}{2}\mathcal{N}(0,1) + \frac{1}{2}\mathcal{N}(3,1)$. The OVL and VUS reach values of 1 and 1/6, respectively.

In Table \ref{Tabla 11}, the rejection proportions are presented. As was the case in previous scenarios under the null hypothesis $H_0: F_1 = F_2 = F_3$, the observed values are generally close to the significance level $\alpha = 0.05$ and $\widehat{OVL}_K$ is conservative.

\begin{table}[H]
\centering
\renewcommand{\arraystretch}{1.25}
\caption{Proportions of rejection for $F_1 \sim \frac{1}{2}\mathcal{N}(0,1) + \frac{1}{2}\mathcal{N}(3,1)$, $F_2 \sim \frac{1}{2}\mathcal{N}(0,1) + \frac{1}{2}\mathcal{N}(3,1)$ and $F_3 \sim \frac{1}{2}\mathcal{N}(0,1) + \frac{1}{2}\mathcal{N}(3,1)$}
\small
\begin{tabular}{|l|l|l|l|l|l|}
\hline
\multicolumn{1}{|c|}{\textbf{$\mathbf{(n_x,n_y,n_z)}$}} & $\mathbf{\widehat{OVL}_N^{BC}}$ & $\mathbf{\widehat{VUS}_N^{BC}} $ & $\mathbf{\widehat{OVL}_K}$ & $\mathbf{\widehat{VUS}_K}$ & $\mathbf{\widehat{VUS}_E} $ \\ \hline
$\mathbf{(20, 20, 20)}$ & 0.051 & 0.045 & 0.036 & 0.053 & 0.053 \\
$\mathbf{(20, 20, 30)}$ & 0.043 & 0.052 & 0.034 & 0.052 & 0.049 \\
$\mathbf{(20, 30, 50)}$ & 0.051 & 0.048 & 0.035 & 0.050 & 0.044 \\
$\mathbf{(30, 50, 50)}$ & 0.038 & 0.054 & 0.030 & 0.052 & 0.046 \\
$\mathbf{(50, 50, 50)}$ & 0.060 & 0.057 & 0.033 & 0.054 & 0.050 \\
$\mathbf{(50, 50, 100)}$ & 0.050 & 0.051 & 0.043 & 0.049 & 0.048 \\ 
$\mathbf{(50, 100, 100)}$ & 0.057 & 0.054 & 0.055 & 0.051 & 0.045 \\
$\mathbf{(100, 100, 100)}$ & 0.038 & 0.037 & 0.042 & 0.035 & 0.036 \\ \hline
\end{tabular}
\label{Tabla 11}
\end{table}

In the second scenario, we assume that class 1 follows the distribution $F_1 \sim \frac{1}{2}\mathcal{N}(0,1) + \frac{1}{2}\mathcal{N}(3,1)$, class 2 is distributed as $F_2 \sim \frac{1}{2}\mathcal{N}(1,1) + \frac{1}{2}\mathcal{N}(4,3/2)$, while class 3 follows $F_3 \sim \frac{1}{2}\mathcal{N}(2,1) + \frac{1}{2}\mathcal{N}(5,2)$. The theoretical overlap measure is $OVL = 0.5807$, while the volume under the ROC surface ($VUS$) is $0.3208$.

Table \ref{Tabla 12} shows that $\widehat{VUS}$ achieves rejection proportions significantly higher than those of $\widehat{OVL}$ for situations with small or moderate sample sizes, although the difference is much smaller when using the parametric estimation. For the larger sample sizes analyzed, the results achieved by both metrics are very similar regardless of the estimation method used, achieving maximum power (unity) for $n_x = 100, n_y = 100, n_z = 100$.

\begin{table}[H]
\centering
\renewcommand{\arraystretch}{1.25}
\caption{Proportions of rejection for $F_1 \sim \frac{1}{2}\mathcal{N}(0,1) + \frac{1}{2}\mathcal{N}(3,1)$, $F_2 \sim \frac{1}{2}\mathcal{N}(1,1) + \frac{1}{2}\mathcal{N}(4,3/2)$ and $F_3 \sim \frac{1}{2}\mathcal{N}(2,1) + \frac{1}{2}\mathcal{N}(5,2)$}
\small
\begin{tabular}{|l|l|l|l|l|l|}
\hline
\multicolumn{1}{|c|}{\textbf{$\mathbf{(n_x,n_y,n_z)}$}} & $\mathbf{\widehat{OVL}_N^{BC}}$ & $\mathbf{\widehat{VUS}_N^{BC}} $ & $\mathbf{\widehat{OVL}_K}$ & $\mathbf{\widehat{VUS}_K}$ & $\mathbf{\widehat{VUS}_E} $ \\ \hline
$\mathbf{(20, 20, 20)}$ & 0.750 & \textbf{0.918} & 0.262 & 0.836 & 0.833 \\
$\mathbf{(20, 20, 30)}$ & 0.778 & \textbf{0.943} & 0.305 & 0.857 & 0.863 \\
$\mathbf{(20, 30, 50)}$ & 0.888 & \textbf{0.971} & 0.475 & 0.920 & 0.912 \\
$\mathbf{(30, 50, 50)}$ & 0.971 & \textbf{0.998} & 0.747 & 0.976 & 0.975 \\
$\mathbf{(50, 50, 50)}$ & 0.995 & \textbf{1.000} & 0.890 & 0.994 & 0.993 \\
$\mathbf{(50, 50, 100)}$ & \textbf{1.000} & \textbf{1.000} & 0.962 & \textbf{1.000} & \textbf{1.000} \\ 
$\mathbf{(50, 100, 100)}$ & \textbf{1.000} & \textbf{1.000} & 0.993 & \textbf{1.000} & 0.999 \\
$\mathbf{(100, 100, 100)}$ & \textbf{1.000} & \textbf{1.000} & \textbf{1.000} & \textbf{1.000} & \textbf{1.000} \\ \hline
\end{tabular}
\label{Tabla 12}
\end{table}

We now consider the last scenario of Normal composition, where the random observations composing each class come from populations distributed as follows: $F_1 \sim \frac{1}{2}\mathcal{N}(0,1) + \frac{1}{2}\mathcal{N}(1,1/2)$, $F_2 \sim \frac{1}{2}\mathcal{N}(0,3/2) + \frac{1}{2}\mathcal{N}(1,1)$, and $F_3 \sim \frac{1}{2}\mathcal{N}(0,2) + \frac{1}{2}\mathcal{N}(1,3/2)$. The values obtained for OVL and VUS in this scenario are, respectively, $0.6784$ and $0.1720$.

Table \ref{Tabla 13} reflects significant changes in the rejection proportions compared to Table \ref{Tabla 12}. $\widehat{OVL}$ achieves a clearly superior power to $\widehat{VUS}$ for all sample sizes considered, which does not adequately evaluate the discriminatory power, as the obtained proportions are not too different from what would be expected under the null hypothesis (significance level $\alpha$). And the best, $\widehat{OVL}_N^{BC}$.

\begin{table}[H]
\centering
\renewcommand{\arraystretch}{1.25}
\caption{Proportions of rejection for $F_1 \sim \frac{1}{2}\mathcal{N}(0,1) + \frac{1}{2}\mathcal{N}(1,1/2)$, $F_2 \sim \frac{1}{2}\mathcal{N}(0,3/2) + \frac{1}{2}\mathcal{N}(1,1)$ and $F_3 \sim \frac{1}{2}\mathcal{N}(0,2) + \frac{1}{2}\mathcal{N}(1,3/2)$}
\small
\begin{tabular}{|l|l|l|l|l|l|}
\hline
\multicolumn{1}{|c|}{\textbf{$\mathbf{(n_x,n_y,n_z)}$}} & $\mathbf{\widehat{OVL}_N^{BC}}$ & $\mathbf{\widehat{VUS}_N^{BC}} $ & $\mathbf{\widehat{OVL}_K}$ & $\mathbf{\widehat{VUS}_K}$ & $\mathbf{\widehat{VUS}_E} $ \\ \hline
$\mathbf{(20, 20, 20)}$ & \textbf{0.433} & 0.068 & 0.318 & 0.056 & 0.061 \\
$\mathbf{(20, 20, 30)}$ & \textbf{0.524} & 0.070 & 0.356 & 0.052 & 0.068 \\
$\mathbf{(20, 30, 50)}$ & \textbf{0.656} & 0.053 & 0.494 & 0.045 & 0.047 \\
$\mathbf{(30, 50, 50)}$ & \textbf{0.885} & 0.075 & 0.734 & 0.061 & 0.063 \\
$\mathbf{(50, 50, 50)}$ & \textbf{0.940} & 0.105 & 0.834 & 0.070 & 0.074 \\
$\mathbf{(50, 50, 100)}$ & \textbf{0.986} & 0.071 & 0.911 & 0.054 & 0.056 \\ 
$\mathbf{(50, 100, 100)}$ & \textbf{0.994} & 0.083 & 0.958 & 0.061 & 0.061 \\
$\mathbf{(100, 100, 100)}$ & \textbf{1.000} & 0.117 & 0.999 & 0.079 & 0.078 \\ \hline
\end{tabular}
\label{Tabla 13}
\end{table}

Now, let's consider a new scenario, this time for the composition of Gamma distributions. Suppose that the three classes are identically distributed as $\frac{1}{2}\Gamma(1,1) + \frac{1}{2}\Gamma(4,1)$. The OVL and VUS reach values of $1$ and $1/6$, respectively.

Table \ref{Tabla 14} presents the rejection proportions for this scenario under the null hypothesis $H_0: F_1=F_2=F_3$.

\begin{table}[H]
\centering
\renewcommand{\arraystretch}{1.25}
\caption{Proportions of rejection for $F_1 \sim \frac{1}{2}\Gamma(1,1) + \frac{1}{2}\Gamma(4,1)$, $F_2 \sim \frac{1}{2}\Gamma(1,1) + \frac{1}{2}\Gamma(4,1)$ and $F_3 \sim \frac{1}{2}\Gamma(1,1) + \frac{1}{2}\Gamma(4,1)$}
\small
\begin{tabular}{|l|l|l|l|l|l|l|}
\hline
\multicolumn{1}{|c|}{\textbf{$\mathbf{(n_x,n_y,n_z)}$}} & $\mathbf{\widehat{OVL}_N^{BC}}$ & $\mathbf{\widehat{VUS}_N^{BC}} $ & $\mathbf{\widehat{OVL}_K}$ & $\mathbf{\widehat{VUS}_K}$ & $\mathbf{\widehat{VUS}_E} $ \\ \hline
$\mathbf{(20, 20, 20)}$ & 0.059 & 0.050 & 0.033 & 0.043 & 0.053  \\
$\mathbf{(20, 20, 30)}$ & 0.045 & 0.050 & 0.034 & 0.055 & 0.059 \\
$\mathbf{(20, 30, 50)}$ & 0.047 & 0.055 & 0.037 & 0.054 & 0.058 \\
$\mathbf{(30, 50, 50)}$ & 0.042 & 0.045 & 0.037 & 0.046 & 0.048 \\
$\mathbf{(50, 50, 50)}$ & 0.059 & 0.061 & 0.038 & 0.052 & 0.052 \\
$\mathbf{(50, 50, 100)}$ & 0.044 & 0.040 & 0.032 & 0.045 & 0.044 \\ 
$\mathbf{(50, 100, 100)}$ & 0.058 & 0.056 & 0.059 & 0.054 & 0.053 \\
$\mathbf{(100, 100, 100)}$ & 0.056 & 0.064 & 0.047 & 0.060 & 0.063 \\ \hline
\end{tabular}
\label{Tabla 14}
\end{table}

Now, suppose that class 1 is distributed as $F_1 \sim \frac{1}{2}\Gamma(1,1) + \frac{1}{2}\Gamma(4,1)$, class 2 is distributed as $F_2 \sim \frac{1}{2}\Gamma(2,1) + \frac{1}{2}\Gamma(5,2/3)$, and class 3 is distributed as $F_3 \sim \frac{1}{2}\Gamma(3,1) + \frac{1}{2}\Gamma(6,1/2)$.
As in the previous examples, we calculate the theoretical values of the two measures: $OVL=0.6609$ and $VUS=0.2583$.

Table \ref{Tabla 15} shows that $\widehat{OVL}_N^{BC}$ again achieves the highest power.

\begin{table}[H]
\centering
\renewcommand{\arraystretch}{1.25}
\caption{Proportions of rejection for $F_1 \sim \frac{1}{2}\Gamma(1,1) + \frac{1}{2}\Gamma(4,1)$, $F_2 \sim \frac{1}{2}\Gamma(2,1) + \frac{1}{2}\Gamma(5,2/3)$ and $F_3 \sim \frac{1}{2}\Gamma(3,1) + \frac{1}{2}\Gamma(6,1/2)$}
\small
\begin{tabular}{|l|l|l|l|l|l|}
\hline
\multicolumn{1}{|c|}{\textbf{$\mathbf{(n_x,n_y,n_z)}$}} & $\mathbf{\widehat{OVL}_N^{BC}}$ & $\mathbf{\widehat{VUS}_N^{BC}} $ & $\mathbf{\widehat{OVL}_K}$ & $\mathbf{\widehat{VUS}_K}$ & $\mathbf{\widehat{VUS}_E} $ \\ \hline
$\mathbf{(20, 20, 20)}$ & \textbf{0.619} & 0.583 & 0.238 & 0.407 & 0.491 \\
$\mathbf{(20, 20, 30)}$ & \textbf{0.640} & 0.614 & 0.313 & 0.439 & 0.531 \\
$\mathbf{(20, 30, 50)}$ & \textbf{0.771} & 0.704 & 0.464 & 0.533 & 0.621 \\
$\mathbf{(30, 50, 50)}$ & \textbf{0.915} & 0.780 & 0.652 & 0.609 & 0.725 \\
$\mathbf{(50, 50, 50)}$ & \textbf{0.973} & 0.864 & 0.784 & 0.712 & 0.799 \\
$\mathbf{(50, 50, 100)}$ & \textbf{0.988} & 0.904 & 0.879 & 0.778 & 0.868 \\ 
$\mathbf{(50, 100, 100)}$ & \textbf{0.999} & 0.940 & 0.959 & 0.833 & 0.892 \\
$\mathbf{(100, 100, 100)}$ & \textbf{1.000} & 0.986 & 0.994 & 0.947 & 0.969 \\ \hline
\end{tabular}
\label{Tabla 15}
\end{table}

Now, we proceed to analyze the possibility that the three classes are distributed as a mixture of Normal and Gamma distributions, following an identical distribution $\frac{1}{2}\mathcal{N}(0,1) + \frac{1}{2}\Gamma(4,1)$. The OVL and VUS reach values of $1$ and $1/6$, respectively, as in the rest of the scenarios under the null hypothesis. The rejection proportions under the null hypothesis $H_0: F_1=F_2=F_3$ are shown in Table \ref{Tabla 16}.

\begin{table}[H]
\centering
\renewcommand{\arraystretch}{1.25}
\caption{Proportions of rejection for $\frac{1}{2}\mathcal{N}(0,1) + \frac{1}{2}\Gamma(4,1)$, $F_2 \sim \frac{1}{2}\mathcal{N}(0,1) + \frac{1}{2}\Gamma(4,1)$ and $F_3 \sim \frac{1}{2}\mathcal{N}(0,1) + \frac{1}{2}\Gamma(4,1)$}
\small
\begin{tabular}{|l|l|l|l|l|l|}
\hline
\multicolumn{1}{|c|}{\textbf{$\mathbf{(n_x,n_y,n_z)}$}} & $\mathbf{\widehat{OVL}_N^{BC}}$ & $\mathbf{\widehat{VUS}_N^{BC}} $ & $\mathbf{\widehat{OVL}_K}$ & $\mathbf{\widehat{VUS}_K}$ & $\mathbf{\widehat{VUS}_E} $ \\ \hline
$\mathbf{(20, 20, 20)}$ & 0.059 & 0.058 & 0.039 & 0.057 & 0.053 \\
$\mathbf{(20, 20, 30)}$ & 0.039 & 0.053 & 0.035 & 0.049 & 0.048 \\
$\mathbf{(20, 30, 50)}$ & 0.045 & 0.047 & 0.036 & 0.050 & 0.049 \\
$\mathbf{(30, 50, 50)}$ & 0.049 & 0.058 & 0.042 & 0.061 & 0.065 \\
$\mathbf{(50, 50, 50)}$ & 0.048 & 0.054 & 0.032 & 0.058 & 0.053 \\
$\mathbf{(50, 50, 100)}$ & 0.051 & 0.048 & 0.035 & 0.047 & 0.048 \\ 
$\mathbf{(50, 100, 100)}$ & 0.050 & 0.053 & 0.038 & 0.056 & 0.057 \\
$\mathbf{(100, 100, 100)}$ & 0.047 & 0.052 & 0.035 & 0.053 & 0.054 \\ \hline
\end{tabular}
\label{Tabla 16}
\end{table}

Finally, we consider the distribution function associated with class 1 as $F_1 \sim \frac{1}{2}\mathcal{N}(0,1) + \frac{1}{2}\Gamma(4,1)$, class 2 as $F_2 \sim \frac{1}{2}\mathcal{N}(1,1) + \frac{1}{2}\Gamma(5,2/3)$, and class 3 as $F_3 \sim \frac{1}{2}\mathcal{N}(2,1) + \frac{1}{2}\Gamma(6,1/2)$. Theoretically, the values for the two test statistics considered are $OVL=0.5450$ and $VUS=0.2580$.

From the rejection proportions presented in Table \ref{Tabla 17}, it can be concluded that the test achieving the highest power is the $\widehat{OVL}$ for all 8 sample sizes proposed in the simulation and for all estimation methods used. And the best, $\widehat{OVL}_N^{BC}$.

\begin{table}[H]
\centering
\renewcommand{\arraystretch}{1.25}
\caption{Proportions of rejection for $F_1 \sim \frac{1}{2}\mathcal{N}(0,1) + \frac{1}{2}\Gamma(4,1)$, $F_2 \sim \frac{1}{2}\mathcal{N}(1,1) + \frac{1}{2}\Gamma(5,2/3)$ and $F_3 \sim \frac{1}{2}\mathcal{N}(2,1) + \frac{1}{2}\Gamma(6,1/2)$}
\small
\begin{tabular}{|l|l|l|l|l|l|}
\hline
\multicolumn{1}{|c|}{\textbf{$\mathbf{(n_x,n_y,n_z)}$}} & $\mathbf{\widehat{OVL}_N^{BC}}$ & $\mathbf{\widehat{VUS}_N^{BC}}$ & $\mathbf{\widehat{OVL}_K}$ & $\mathbf{\widehat{VUS}_K}$ & $\mathbf{\widehat{VUS}_E} $ \\ \hline
$\mathbf{(20, 20, 20)}$ & \textbf{0.786} & 0.463 & 0.647 & 0.398 & 0.457 \\
$\mathbf{(20, 20, 30)}$ & \textbf{0.862} & 0.519 & 0.757 & 0.449 & 0.524 \\
$\mathbf{(20, 30, 50)}$ & \textbf{0.943} & 0.598 & 0.892 & 0.541 & 0.616 \\
$\mathbf{(30, 50, 50)}$ & \textbf{0.993} & 0.680 & 0.987 & 0.613 & 0.692 \\
$\mathbf{(50, 50, 50)}$ & \textbf{1.000} & 0.755 & 0.993 & 0.724 & 0.783 \\
$\mathbf{(50, 50, 100)}$ & \textbf{1.000} & 0.811 & 0.999 & 0.775 & 0.847 \\ 
$\mathbf{(50, 100, 100)}$ & \textbf{1.000} & 0.879 & \textbf{1.000} & 0.859 & 0.895 \\
$\mathbf{(100, 100, 100)}$ & \textbf{1.000} & 0.946 & \textbf{1.000} & 0.938 & 0.960 \\ \hline
\end{tabular}
\label{Tabla 17}
\end{table}

\section{Concluding Remarks}

The volume under the ROC surface (VUS) can be regarded as an important generalization of the area under a ROC curve to accommodate problems of three-class classification. It has been widely studied for the assessment of diagnostic markers in this context but not many other measures have been proposed. In this work, we have demonstrated that VUS fails as AUC does in similar situations. For scenarios where the standard deviation varies between distributions associated with different classes, the test based on VUS tends to perform very poorly, with power values approaching to 0.05 (see Tables \ref{Tabla 3}, \ref{Tabla 6} and \ref{Tabla 13}). Therefore, alternative measures to VUS are necessary. \hyperref[Samawi et al. (2017)]{Samawi et al. (2017)} and \hyperref[Pardo and Franco-Pereira (2024)]{Pardo and Franco-Pereira (2024)} proposed an overlap measure and stated its advantages over the AUC for distinguishing between 2 underlying populations for healthy and diseased. We have proposed an overlap measure (OVL) for the case of three diagnostic groups. It results in outperforming VUS in most cases. Only in a few cases VUS has higher power than OVL. Furthermore, in these cases, VUS outperforms slightly OVL but, in some cases, as pointed out before, to use VUS would naively lead to rejection of informative biomarkers.

Parametric and non-parametric estimations of OVL have been proposed. Surprisingly, the parametric approach is better in terms of power than the kernel approach but not only under normal assumptions but under non-normal after applying Box-Cox transformation.

Our measure can have a broad application in the biomarker field as it may be safely applied independently of the underlaid distributions. An application to the evaluation of Alzheimer’s biomarkers illustrates our proposal.

Future research involves the development of accurate methods for confidence intervals construction for OVL in the three-class classification problems as well as to account for possible effects of covariates on the OVL.

\bigskip

\noindent \textbf{Acknowledgments}

This work was partially supported by grant PID2022-137050NB-I00, funded by\\ MCIN/AEI/10.13039/501100011033/ and by ERDF A way of making Europe.

\end{document}